\documentclass[12pt]{JHEP3}

\usepackage{mathrsfs}
\usepackage{amsmath,amssymb}
\usepackage{epsfig}
\usepackage{relsize}
\usepackage{graphicx}

\def\sl(2){\alg{sl}(2)}

\def\be{\begin{equation}}
\def\ee{\end{equation}}

\newcommand{\bea}{\begin{eqnarray}}
\newcommand{\eea}{\end{eqnarray}}

\def\a {\alpha}

\def\pa {\partial}

\def\g {\gamma}

\def\la{\label}
\def\e{\epsilon}
\def\ov{\over}

\def\tp{{\widetilde p}}

\newcommand{\alg}[1]{\mathfrak{#1}}

\newcommand{\AdS}{{\rm  AdS}_5\times {\rm S}^5}
\newcommand{\ads}{{\rm  AdS}_5\times {\rm S}^5}

\newcommand{\bem}{\left (\begin{matrix}}
\newcommand{\eem}{\end{matrix} \right )}

\author{Gleb Arutyunov$^a$\footnote{Email: G.E.Arutyunov@uu.nl, frolovs@maths.tcd.ie} {}\footnote{Correspondent fellow at Steklov
Mathematical Institute, Moscow.}\, and\,  Sergey Frolov$^{b\,
\dagger}$
 \\ $^{a}$ {\it Institute for Theoretical
Physics and Spinoza Institute,\\ ~~Utrecht University, 3508 TD
Utrecht, The Netherlands} \\ $^b$ {\it Hamilton Mathematics Institute and School of Mathematics, \\
~~Trinity College, Dublin 2, Ireland} }

\abstract{ We use the recently found integral representation for
the dressing phase in the kinematic region of the mirror theory to
simplify the TBA equations for the $\AdS$ mirror model. The resulting
set of equations provides an efficient starting point for both
analytic and numerical studies.}

\title{Simplified TBA equations of\\
the $\AdS$ mirror model}

\preprint{
          \smaller{\smaller{\smaller{ITP-UU-09-28}}}\\[-.5ex]
          \smaller{\smaller{\smaller{SPIN-09-25}}}\\[-.5ex]
          \smaller{\smaller{\smaller{TCDMATH 09-17}}}\\[-.5ex]
          \smaller{\smaller{\smaller{HMI-09-08}}}}

\begin{document}

\renewcommand{\thefootnote}{\arabic{footnote}}
\setcounter{footnote}{0}


\section{Introduction}
Recently there has been a substantial progress towards solving the
finite-size spectral problem of the AdS/CFT correspondence
\cite{M}. First, a perturbative approach due to L\"uscher  has
been generalized to the case of the non-Lorentz invariant
light-cone string sigma model on $\ads$ and further applied to
find the four- and five-loop anomalous dimensions of the Konishi
operator \cite{BJ,BJLH}; the four-loop result exhibits a
remarkable agreement with the direct field-theoretic computation
\cite{Sieg,Vel}. Second, the TBA approach \cite{za} based on the
so-called $\ads$ mirror model \cite{AFtba} has been advanced as a
mean to determine the {\it exact} string spectrum.\footnote{The
importance of the TBA approach in the AdS/CFT spectral problem was
stressed in \cite{AJK} where it was used to explain wrapping
effects in gauge theory.} In particular, the TBA equations for the
ground state of the light-cone superstring were
derived in \cite{AFsh}-\cite{GKV2}. Another important tool for studying the finite-size
spectral problem, namely, the so-called Y-system has been proposed
in \cite{GKV1}, and its general solution has been constructed in
\cite{Heg}. Upon specifying an analytic behavior, solutions of the
Y-system should also describe  the excited states of the model.
Comparison of the TBA equations to those of the Y-system
\cite{AFmtba} reveals intricate analytic properties of the latter
indicating that, in contrast to relativistic models, the
corresponding Y-system should be defined on an infinite genus
Riemann surface \cite{FS}. Finally, we point out that in the work
\cite{GKV2} integral (TBA-like) equations for excited states in
the $\sl(2)$ sector have been suggested along the lines of
\cite{BLZe,DT,Teschner:2007ng,BJ} and they were further used in
\cite{GKV3} to compute numerically an all-loop anomalous dimension
supposedly corresponding to one of the descendents of the Konishi
operator. The subleading at strong coupling $\lambda^{-1/4}$-term
found from this computation disagrees however with the result by
\cite{RT} and the origin of this disagreement remains unclear for
the moment. Some important subtleties concerning the non-analytic
behavior of the asymptotic string energies at strong coupling have
been pointed out in the recent work \cite{SR}.

Needless to say that the TBA/Y-system equations proposed above
have a number of unusual features which call for a deeper
understanding of their structure and analytic properties. In this
note we will make a further step in simplifying the TBA equations
which follow from the corresponding string hypothesis \cite{AFsh}
by using the canonical procedure \cite{Korepin}. We recall that
the Y-system is obtained from the {\it canonical} TBA equations by
acting on the latter with the discrete Laplace operator
$\Delta_{MN}$, where $M,M=1,\ldots,\infty$. The Laplace operator
has the following representation $\Delta_{MN}=(K+1)_{MN}^{-1}\star
s^{-1}$, where $K+1$ is a certain invertible operator and $s^{-1}$
is an operator which has a null space, {\it i.e.} in general
$f\star s^{-1}\star s\neq f$. The fist simplification of the
canonical TBA system occurs when acting on it with $K+1$, because
it brings most of the TBA equations to the local form, see
\cite{AFmtba} for details.

There are further simplifications we point out in this note. The
first one concerns the infinite sums involving the Y-function
$Y_{M|vw}$ and $Y_{M|w}$ for the so-called $M|vw$- and
$M|w$-strings \cite{AFmtba}. These infinite sums occur in some of
the TBA equations and they are difficult for numerical studies due
to their rather slow convergence properties. We show that by using
certain identities between the TBA kernels these sums can be
removed in favor of infinite sums involving $Y_Q$-functions only,
the latter have much better convergence properties.

The second  simplification concerns the main TBA equation for the
$Q$-particles which involves the contribution of the dressing
phase \cite{AFS}. This phase is nothing else but the BES
expression \cite{BES} analytically continued to the kinematic
region of the mirror theory \cite{AFtba}. In \cite{AFdf} we have
obtained a convenient integral representation for this analytic
continuation starting from the DHM representation \cite{DHM} valid
in the kinematic region of the original string theory. Here we
will work out explicitly the action of the operator $(K+1)^{-1}$
on the mirror dressing phase given by this integral representation
and find a very simple final expression.

We believe that the simplification procedure developed here can
also be applied to the integral equations describing the excited
states, although there are new important subtleties related to
singularities of certain Y-functions that should be taken into
account.

The note is organized as follows. In the next section we present
the main result on the simplified TBA equations. The interested
reader can find the details of our derivation in two appendices.

\newpage

\section{Simplified TBA equations}\la{sect:tba1}

We recall \cite{AFmtba} that the spectrum of the mirror model in
the thermodynamic limit contains $Q$-particles with pseudo-energy
$\e_{Q}$, two copies of $M|w$- and $M|vw$-strings with
pseudo-energies $\e_{M|w}^{(\a)}$ and $\e_{M|vw}^{(\a)}$, where
$\a=1,2$, and, finally, two copies of $y^{\pm}$-particles, whose
pseudo-energies $\e_{y^\pm}^{(\a)}$ are supported on the interval
$[-2,2]$ of the rapidity variable $u$. The pseudo-energies and
densities for all the other particles are defined for all real
values of $u$. It is convenient to introduce  so-called Y-functions which are
related to the pseudo-energies as
\bea\la{Yfa} Y_{Q} =
e^{-\e_{Q}}\,,\quad Y_{M|vw}^{(\a)} = e^{\e_{M|vw}^{(\a)}}\,,\quad
Y_{M|w}^{(\a)} = e^{\e_{M|w}^{(\a)}}\,,\quad Y_\pm^{(\a)}
=e^{\e_{y^\pm}^{(\a)}}\,,\quad \a=1,2\,. \eea

By using the integral representation for the mirror model dressing factor \cite{AFdf}, the
partially simplified set of the TBA equations obtained in \cite{AFmtba}
can be brought to the form
\begin{itemize}
\item $M|w$-strings: $\ M\ge 1\ $, $Y_{0|w}^{(\a)}=0$
\bea\la{Yforw} \log Y_{M|w}^{(\a)}=  \log(1 +  Y_{M-1|w}^{(\a)}
)(1 +  Y_{M+1|w}^{(\a)} )\star s +\delta_{M1}\,
\log{1-{e^{ih_\a}\ov Y_-^{(\a)}}\ov 1-{e^{ih_\a}\ov Y_+^{(\a)}}
}\star s\,~~~~~ \eea

\item $M|vw$-strings: $\ M\ge 1\ $, $Y_{0|vw}^{(\a)}=0$
\bea\la{Yforvw} \hspace{-0.3cm}\log Y_{M|vw}^{(\a)}&=&\log(1 +
Y_{M-1|vw}^{(\a)} )(1 +  Y_{M+1|vw}^{(\a)} )\star s\\\nonumber &-&
\log(1 +  Y_{M+1})\star s+\delta_{M1}
\log{1-e^{-ih_\a}Y_-^{(\a)}\ov 1-e^{-ih_\a}Y_+^{(\a)}}\star
s\,~~~~~ \eea

\item $y$-particles
\bea
\la{Yfory} &&\log Y_\pm^{(\a)} =  - \log\left(1+Y_Q
\right)\star K_\pm^{Qy}+\log {1+{1\ov Y_{M|vw}^{(\a)}}
 \ov1+{1\ov Y_{M|w}^{(\a)}}}\star K_M\,
\eea

\item $Q$-particles for $Q\ge 2$
\bea \log Y_{Q}&=&\log{ \left(1 +  {1\ov Y_{Q-1|vw}^{(1)}}
\right)\left(1 +  {1\ov Y_{Q-1|vw}^{(2)}} \right)\ov (1 +
{1\ov Y_{Q-1}} )(1 + {1\ov Y_{Q+1}} ) }\star
s\la{YforQ2}  \,~~~~~~~ \eea

\item $Q=1$-particle
\bea
\la{YforQ1}
\log Y_{1}&=&
\log{\left(1-{e^{ih_1} \ov Y_{-}^{(1)}} \right)\left(1-{e^{ih_2} \ov Y_{-}^{(2)}}\right)\ov 1 +  {1\ov Y_{2}} }\star s - \check\Delta\star s \,,~~~~~ \eea
where
\bea\la{addit}
\check\Delta&=& L\, \check{\cal E} +\log\left(1-{e^{ih_1} \ov Y_{-}^{(1)}}
\right)\left(1-{e^{ih_2} \ov Y_{-}^{(2)}}\right) \left(1-{e^{ih_1}
\ov Y_{+}^{(1)}} \right)\left(1-{e^{ih_2} \ov Y_{+}^{(2)}}\right)
\star \check{K}~~~~~~~~~
\\
\nonumber&+& \log\left(1 +  {1\ov Y_{M|vw}^{(1)}} \right)\left(1 +
{1\ov Y_{M|vw}^{(2)}} \right)\star \check{K}_M +2\log\left(1+Y_{Q}
\right)\star {\check K}^\Sigma_{Q}  \,.~~~~~ \eea

\end{itemize}
Let us stress that in the convolutions involving $Y_\pm^{(\a)}$-functions one has to integrate over the interval $[-2,2]$.
In eq.\eqref{addit} $L$ coincides with the light-cone momentum
$P_+$ of the $\AdS$ string theory in the light-cone gauge, which
is simultaneously the circumference of a cylinder on which the
corresponding string sigma model is defined. In this paper we
consider only the $a=0$ light-cone gauge (or  temporal gauge)
\cite{AF04,AFrev} where $L=J$, and $J$ is one of the ${\rm SO}
(6)$ charges carried by  the string. Also, $h_{\a}=(-1)^{\a}h$,
where $h$ can be thought of as the chemical potential for
fermionic particles.

The energy of the ground state of the light-cone gauge-fixed
string theory on $\AdS$ defined on a cylinder of circumference $L$
is expressed through the Y-functions which solve the TBA equations as follows
 \bea
\la{energyLa} E_h(L) &=&-\int {\rm d}u\, \sum_{Q=1}^\infty{1\ov
2\pi}{d\tp_Q\ov du}\log\left(1+Y_Q\right)\,. \eea

Equations above involve convolutions with a number of kernels
which we specify in appendix \ref{app:kern}.

The TBA equations \eqref{Yfory}  for $y_\pm$-particles contain the
infinite sum involving the Y-functions for $vw$- and $w$-strings.
It can be replaced by a sum of terms which only involve
$Y_{1|vw}^{(\a)}$, $Y_{1|w}^{(\a)}$ and $Y_Q$ by using the
following formula derived in \cite{AFmtba}
 \bea\la{Yforvww} \log{1 +  {1\ov
Y_{M|vw}^{(\a)}} \ov 1 +  {1\ov Y_{M|w}^{(\a)}} } \star K_M&=&
\log{1 +  Y_{1|vw}^{(\a)} \ov 1 +  Y_{1|w}^{(\a)} }  \star
s\\\nonumber
 &+& \log(1 +  Y_{Q+1}) \star s\star K_Q+  \log(1 +  Y_{Q})\star K_{Qy} \star s\star K_1\,.~~~~~
\eea Since $Y_Q$-functions decrease very fast for large $Q$, the
formula  (\ref{Yforvww}) seems to be useful for numerical studies
of the TBA equations.  Let us also mention that
 the last sum in eq.(\ref{Yforvww}) can be expressed in terms of $Y_\pm$-functions by using the
 formula that follows from eq.(\ref{Yfory})
\bea\la{YforQypm}
 \log(1 +  Y_{Q})\star K_{Qy}= \log{Y_+^{(\a)}\ov Y_-^{(\a)}} = {1\ov 2}\log\frac{Y_+^{(1)}}{Y_-^{(1)}}\frac{Y_+^{(2)}}{Y_-^{(2)}} \,,\qquad \a=1,2\,.~~~~~
\eea

The kernel ${\check K}^\Sigma_{Q}$ corresponding to the improved
dressing factor is worked out in appendix B. It has the following
representation
$$
{\check K}^\Sigma_{Q} = - K_{Qy}\star  \check{I}_0 + \check{I}_Q\,
,
$$
where $\check{I}_Q$ is given by eq.(\ref{IQ}). Due to this
representation of ${\check K}^\Sigma_{Q}$, the formula
\eqref{YforQypm} can be also used to partially exclude the
infinite contribution of $Q$-particles in eq.(\ref{addit}) in
favor of the $y$-particles.

Finally, $\check\Delta$ in the TBA equation \eqref{YforQ1} contains
another infinite sum involving the Y-functions for $vw$-strings.
This sum can be  expressed in terms of $Y_Q$-functions only by
using  the following identity that holds outside the interval
$[-2,2]$ \bea\la{YvwYvw} \log\left(1 +  {1\ov Y_{M|vw}^{(1)}}
\right)\left(1 + {1\ov Y_{M|vw}^{(2)}} \right)\star \check{K}_M
&=&
\log\left(1+Y_Q\right)\left(1+Y_{Q+2}\right)\star\check{K}_Q\\\nonumber
&+&\log Y_2 +2 \log Y_2\star \check{K} -\log{Y_1}\star \check{K}_1
\,, \eea where both sums in the left and the right hand side are
from 1 to $\infty$. Note also
that at large $L$ the r.h.s. of \eqref{YvwYvw} is finite because $\widetilde{\cal E}_2 +2 \widetilde{\cal E}_2\star \check{K} -\widetilde{\cal E}_1\star \check{K}_1 =0$, and we recall that in the term $ \widetilde{\cal E}_2\star \check{K}$ one integrates over the interval $[-2,2]$.
In fact for the ground state the l.h.s. of \eqref{YvwYvw} goes to 0  in the large $L$ limit \cite{FS}, and it can be also easily seen from the r.h.s. of \eqref{YvwYvw} by using that
$1\star  \check{K} = -{1\ov 2}\big( \theta(-u-2)+ \theta(u+2)\big)$ and $1\star  \check{K}_M = 0$.

To prove this formula, one should use eq.\eqref{YforQ2}, and the
following identities which hold for $|v|>2$ \bea
&&\hspace{-0.4cm}  s\star \left(\check{K}_{Q-1} + \check{K}_{Q+1} \right) = \check{K}_{Q}\,,\quad Q=2,3,\cdots,\infty\,,\\
&&\hspace{-0.4cm} s\star \check{K}_{2} = \check{K}_{1} - s -
2s\star \check{K}= \check{K}_{1}(u,v) - s(u-v) - 2\int_{-2}^2\,
dt\, s(u-t) \check{K}(t-v)\,.~~~~~~~~ \eea

As the result of these simplifications the quantity $\check\Delta$ in
eq.(\ref{addit}) can be written in the form
 \bea\la{addit2} \check\Delta&=& L\, \check{\cal E}
+\log\left(1-{e^{ih_1} \ov Y_{-}^{(1)}}
\right)\left(1-{e^{ih_2} \ov Y_{-}^{(2)}}\right) \left(1-{e^{ih_1}
\ov Y_{+}^{(1)}} \right)\left(1-{e^{ih_2} \ov Y_{+}^{(2)}}\right)
\star \check{K}
\\
\nonumber&+&
\log Y_2 +2 \log Y_2\star \check{K} -\log{Y_1}\star \check{K}_1
+\log\left(1+Y_{Q} \right)\star\big( 2\check{K}_Q^\Sigma +
\check{K}_Q  +\check{K}_{Q-2}\big)  \,. \eea To recall, in the
last formula the sums over $Q$ run from $1$ to $\infty$,  the
convolutions involving  $\check{K}$ are taken over the interval
$[-2,2]$, and we use the convention
$\check{K}_{-1}=\check{K}_{0}=0$.

Thus, eqs.\eqref{Yforvww} and \eqref{YvwYvw} allow one to exclude
from all the TBA equations the infinite sums  involving the
functions $Y_{M|vw}^{(\a)}$ and $Y_{M|w}^{(\a)}$. The resulting
set of equations is well-suited for both analytic and numerical
studies. In particular, one could analyze the behavior of the
energy \eqref{energyLa} as a function of the complexified length
$L$ or coupling constant $g$.

\section*{Acknowledgements}
The work of G.~A. was supported in part by the RFBR grant
08-01-00281-a, by the grant NSh-672.2006.1, by NWO grant 047017015
and by the INTAS contract 03-51-6346. The work of S.F. was
supported in part by the Science Foundation Ireland under Grants
No. 07/RFP/PHYF104 and 09/RFP/PHY2142.

\section{Appendix A. Kernels}\la{app:kern}

All kernels and S-matrices we are using are expressed in terms of the function
$x(u)$
 \bea\la{basicx}
x(u)=\frac{1}{2}(u-i\sqrt{4-u^2}), ~~~~{\rm Im}\, x(u)<0\, , \eea
which maps the $u$-plane with the cuts $[-\infty, -2]\cup [2,\infty]$ onto the physical region of the mirror theory,
and the function $x_s(u)$
\bea\la{stringx}
x_s(u)={u\ov 2}\Big(1 + \sqrt{1-{4\ov u^2}}\Big)\,,\quad |x_s(u)|\ge 1\,,
\eea
which maps the $u$-plane with the cut $[-2,2]$ onto the physical region of the string theory.

The momentum $\tilde{p}^Q$ and the energy $\tilde{\cal{E}}_Q$ of a
mirror $Q$-particle are expressed in terms of $x(u)$ as follows
\bea \tp_Q=g x\big(u-\frac{i}{g}Q\big)-g
x\big(u+\frac{i}{g}Q\big)+i Q\, ,
~~~~~\tilde{\cal{E}}_Q=\log\frac{x\big(u-\frac{i}{g}Q\big)}{x\big(u+\frac{i}{g}Q\big)}\,
. \eea

The TBA equations discussed in section \ref{sect:tba1}  involve convolutions with a number of kernels
which we specify below, see also \cite{AFmtba} for more details
and the definition of the convolutions. First,  all the TBA equations contain the following universal kernels
\begin{alignat}{2}
s (u) & = \frac{1}{2 \pi i} \, \frac{d}{du} \log S(u)= {g \ov 4\cosh {\pi g u \ov 2}}\,,\quad S(u)=\tanh[ \frac{\pi}{4}(u g - i)]\,,
\label{s-kern} \\
K_Q (u) &= \frac{1}{2\pi i} \, \frac{d}{du} \, \log S_Q(u) = \frac{1}{\pi} \, \frac{g\, Q}{Q^2 + g^2 u^2}\,,\quad S_Q(u)= \frac{u - \frac{iQ}{g}}{u + \frac{i Q}{g}} \,, \label{KMu}
\end{alignat}
which appear in TBA equations of any integrable model. Note that the kernel $K_{Q}$ has an
interesting group property
$$
K_Q\star K_{Q'}=K_{Q'}\star K_{Q}=K_{Q+Q'}\, ,
$$
where the integrals in the convolution are taken from $-\infty$ to
$+\infty$.

Then, the kernels $K_\pm^{Qy}$ are related to the scattering matrices $S_\pm^{Qy}$ of $Q$- and $y_\pm$-particles in the usual way
\bea\nonumber
K^{Qy}_-(u,v)&=&1\ov 2\pi i}{d\ov du}\log S^{Qy}_-(u,v)\,,\quad  S^{Qy}_-(u,v) = \frac{x(u-i{Q\ov g})-x(v)}{x(u+i{Q\ov g})-x(v)} \sqrt{{\frac{x(u+i{Q\ov g})}{x(u-i{Q\ov g})}} \,,\\\nonumber
K^{Qy}_+(u,v)&=&1\ov 2\pi i}{d\ov du}\log S^{Qy}_+(u,v)\,,\quad  S^{Qy}_+(u,v) =  \frac{x(u-i{Q\ov g})-{1\ov x(v)}}{x(u+i{Q\ov g})-{1\ov x(v)}} \sqrt{{\frac{x(u+i{Q\ov g})}{x(u-i{Q\ov g})}}\,.\\\la{KpmQy} \eea
These kernels can be expressed in terms of the kernel $K_Q$, and the  kernel
 \bea
 K(u,v) = \frac{1}{2 \pi i} \, \frac{d}{du} \, \log\Big( \frac{x(u) - x(v)}{x(u) - 1/x(v)}\Big) = \frac{1}{2 \pi i} \, \frac{ \sqrt{4-v^2}}{\sqrt{4-u^2}}\, {1\ov u-v} \,,
\label{Kuv}
 \eea
 as follows
 \bea
 K^{Qy}_\mp(u,v)&=&{1\ov 2}\Big( K_Q(u-v) \pm  K_{Qy}(u,v)\Big)\,,
 \eea
 where  $K_{Qy}$ is given by \bea
K_{Qy}(u,v)=K(u-\frac{i}{g}Q,v)-K(u+\frac{i}{g}Q,v)\, .\eea
Next, we introduce the following kernel
\bea \la{bK}
\bar{K}(u,v)= \frac{1}{2 \pi i} \, \frac{d}{du} \, \log\Big( \frac{x(u) - x_s(v)}{x(u) - 1/x_s(v)}\Big) ={1\ov
2\pi} \frac{\sqrt{1-{4\ov v^2}}}{\sqrt{4-u^2}}{v\ov u-v}\,,
\eea
This kernel \eqref{bK} can be thought of as an analytic continuation of $K(u,v)$
from the mirror theory $v$-plane to the string theory one.
With the help of this kernel we can now define\footnote{The definitions of the kernels $\check{K}$ and $\check{K}_Q$ differ by the sign from the ones used in \cite{AFmtba}.}
 \bea
\check{K}(u,v)&=&\bar{K}(u,v)\big[\theta(-v-2)+\theta(v-2)\big]\,,\quad
~~~~\\
 \la{cKQuva} \check{K}_Q (u,v)&=& \big[\bar{K}(u+{i\ov
g}Q,v) + \bar{K}(u-{i\ov g}Q,v) \big]\big[\theta(-v-2)
+\theta(v-2)\big]\, , \eea where $\theta(u)$ is the standard unit
step function. Obviously, both $\check{K}$ and $\check{K}_Q$
vanish for $v$ being in the interval $(-2,2)$ and are equal to (twice) the jump discontinuity of the kernels ${K}$ and ${K}_{Qy}$ across the real semi-lines $|v|>2$.

The quantity
$\check{\cal E}$ is defined as
\bea\la{cEu}
\check{\cal E}(u)=\log  \frac{x (u - i0)}{x (u + i0)} =2\log |x_s(u)|  \neq 0 \quad {\rm for} \ \ |u|>2 \, .
\eea
Finally, eq.(\ref{YforQ1}) involves the
  kernel
  \bea\la{Ksig}
 {\check K}^\Sigma_{Q} = {1\ov 2\pi i} {\pa\ov \pa u} \log{\check \Sigma}_{Q}= - K_{Qy}\star  \check{I}_0 + \check{I}_Q
  \eea
  where
   \bea\la{checkIQ}
 &&\check{I}_Q=\sum_{n=1}^\infty
\check{K}_{2n+Q}(u,v)=K_\Gamma^{[Q+2]}(u-v)+2\int_{-2}^2 {\rm d}t
\, K_\Gamma^{[Q+2]}(u-t)\check{K}(t,v) \,,~~~~~~~\\\la{KG0}
&&K_\Gamma^{[Q]}(u)={1\ov 2\pi i} {d\ov d u} \log
\frac{\Gamma\big[{Q\ov 2}-\frac{i}{2}g u\big]}{\Gamma\big[{Q\ov
2}+\frac{i}{2}g u\big]}={g\g\ov2\pi}+ \sum_{n=1}^\infty\Big(K_{2n+Q-2}(u)-{g\ov 2\pi n}\Big)  \, .~~~~~~~~~~\eea
The kernel \eqref{Ksig} is related to the dressing kernel
 \bea
 K_{QQ'}^{\Sigma}(u,u')=\frac{1}{2\pi i}\frac{d}{du}\log \Sigma_{QQ'}(u,u')\, ,
\eea
where $\Sigma_{QQ'}(u,v)$ is the improved dressing factor \cite{AFdf}
obtained by fusing the $\sl(2)$ S-matrices for individual
constituents of the bound states in the mirror theory. The relation is given by
\bea\la{KSKi1}
\check{K}_{Q}^\Sigma(u,v)= K^\Sigma_{Q1}(u,v+{i\ov g}-i0)+K^\Sigma_{Q1}(u,v-{i\ov g}+i0) - K^\Sigma_{Q2}(u,v)
\,, \eea
and will be proven in the next section.

\section{Appendix B. Simplifying the dressing kernel contribution}\la{app:drkern}

\subsection{$\Phi$ and $\Psi$ functions}

Below we present the functions $\Phi$ and $\Psi$ used to represent
the dressing phase in the kinematic region of the mirror theory
\bea\la{Phip} \Phi(x_1,x_2)&=&i\oint\frac{{\rm d}w_1}{2\pi i}\oint
\frac{{\rm d}w_2}{2\pi i}\frac{1}{(w_1-x_1)(w_2-x_2)}I(w_1,w_2) \,
, ~~~~
\\
 \la{Psi1} \Psi({x_1},x_2)&=&i\oint\frac{{\rm d }w}{2\pi i}
\frac{1}{w-x_2}I(x_1,w)\, ,~~~~~~~ \eea where
 \bea
I(w_1,w_2)=\log{\Gamma\big[1+{i\ov
2}g\big(w_1+\frac{1}{w_1}-w_2-{1\ov w_2}\big)\big]\ov
\Gamma\big[1-{i\ov 2}g\big(w_1+\frac{1}{w_1}-w_2-{1\ov
w_2}\big)\big]} \eea and the integrals are over the unit circles.
Both functions are discontinuous through the unit circle.
Considering $\Psi(x_1,x_2)$ as a function of the rapidity variable
$u$ through $x_1\equiv x(u)$, where $x(u)$ is given by
eq.(\ref{basicx}), it has an infinite number of cuts located at
$u\pm \frac{2i}{g}n$, $-2\leq u\leq 2$ and $n=1,2,\infty$. Both
$\Phi$ and $\Psi$ are discontinuous when $x_1$ or $x_2$ for $\Phi$
and $x_2$ for $\Psi$ crosses the unit circle. The corresponding
jump discontinuities are given in \cite{AFdf}.

\subsection{Improved dressing factor}

As we have shown in our recent work \cite{AFdf}, the improved
dressing factor in the kinematic region of the mirror theory does
not depend on the internal structure of a bound state employed in
the fusion procedure. Also, a convenient integral representation
for this factor has been found in \cite{AFdf}, namely
\bea\la{sigtot3} \begin{aligned} {1\ov i}\log\Sigma_{QQ'}(y_1,y_2)
 &=
\Phi(y_1^+,y_2^+)-\Phi(y_1^+,y_2^-)-\Phi(y_1^-,y_2^+)+\Phi(y_1^-,y_2^-)
\\ &-{1\ov 2}\left(\Psi(y_1^+,y_2^+)+\Psi(y_1^-,y_2^+)-\Psi(y_1^+,y_2^-)-\Psi(y_1^-,y_2^-)\right) \\
&+{1\ov
2}\left(\Psi(y_{2}^+,y_1^+)+\Psi(y_{2}^-,y_1^+)-\Psi(y_{2}^+,y_1^-)
-\Psi(y_{2}^-,y_1^-) \right)
 \\
&+{1\ov i}\log\frac{ i^{Q}\,\Gamma\big[Q'-{i\ov
2}g\big(y_1^++\frac{1}{y_1^+}-y_2^+-\frac{1}{y_2^+}\big)\big]} {
i^{Q'}\Gamma\big[Q+{i\ov
2}g\big(y_1^++\frac{1}{y_1^+}-y_2^+-\frac{1}{y_2^+}\big)\big]}{1-
{1\ov y_1^+y_2^-}\ov 1-{1\ov
y_1^-y_2^+}}\sqrt{\frac{y_1^+y_2^-}{y_1^-y_2^+}} \,.~~~~~
\end{aligned}\eea
Here $y_{1,2}^{\pm}$ are parameters
of $Q$ and $Q'$-particle bound states in the mirror theory.
The
bound state parameters read as
 \bea
y_1^+=x(u+\frac{i}{g}Q)\, , ~~~~y_1^-=x(u-\frac{i}{g}Q)\, ,\\
y_2^+=x(u'+\frac{i}{g}Q')\, , ~~~~y_2^-=x(u'-\frac{i}{g}Q')\, \eea
In the next subsection we will use this integral representation for
the dressing kernel together with the properties of other kernels
involved in the TBA equation to simplify the dressing kernel contribution to the TBA equation (\ref{YforQ1}) which contains this kernel.

\subsection{Computing $\check{K}_{Q}^\Sigma$ }
In \cite{AFmtba} we have conjectured the following relation
\bea\la{KSKi} K^\Sigma_{QQ'}\star (K + 1)^{-1}_{Q'Q''}
\stackrel{?}{=}\delta_{1Q''}\check{K}_{Q}^\Sigma\star s\,, \eea where the
kernel $\check{K}_{Q}^\Sigma(u,v)$ is supposed to vanish for
$|v|<2$. With an explicit expression (\ref{sigtot3}) for the
dressing kernel at hand, we can now verify this conjecture and
find $\check{K}_{Q}^\Sigma$.

\smallskip

Denote by $\Delta_{Q'Q''}$ the discrete Laplace operator \bea
\Delta_{Q'Q''}\equiv (K+1)^{-1}_{Q'Q''}\star
s^{-1}=\delta_{Q'Q''}s^{-1}-(\delta_{Q'+1,Q''}+\delta_{Q'-1,Q''})\,
. \eea For $Q''=1$ this is not  anymore the Laplace operator, but
we continue to use the same notation, {\it i.e.}, \bea \Delta_{Q'
1}\equiv (K+1)^{-1}_{Q'1}\star s^{-1}=\delta_{Q'\,
1}s^{-1}-\delta_{Q' 2}\, . \eea As was shown in \cite{AFdf}, the
improved dressing factor is a holomorphic function of its
arguments in the intersection of the region
$\{|y^{+}_{1,2}|<1,|y^{-}_{1,2}|>1\}$ with the mirror region ${\rm
Im}\,y_i^{\pm}<0$, which includes the real momentum line of the
mirror theory. This immediately implies that \bea
 K^\Sigma_{QQ'}\star \Delta_{Q'Q''}=0~~~~{\rm for}~~~Q''\neq 1.
\eea

Now we consider $\frac{1}{i}\log \Sigma_{QQ'}\star \Delta_{Q'1}$.
We have to distinguished two cases, $|v|<2$ and $|v|>2$. We start
with the first case.

\vskip 0.4cm \noindent \underline{\bf Case I: $|v|<2$.}

\bigskip

\noindent The formula (\ref{sigtot3}) contains four lines which
contributions we will work out separately. The computation
proceeds as follows \bea
&&\Phi(y_1^+,y_2^+)\star \Delta_{Q'1}=\Phi\Big[y_1^+,x(v+\frac{i}{g}Q')\Big]\star\Delta_{Q'1}=\\
&& \nonumber
\Phi\Big[y_1^+,x(v+\frac{2i}{g}-i0)\Big]+\Phi\Big[y_1^+,x(v+i0)\Big]
-\Phi\Big[y_1^+,x(v+\frac{2i}{g})\Big] =
\Phi\big[y_1^+,x(v+i0)\big]\, . \nonumber\eea Thus, for the
difference of two $\Phi$-functions with the same first argument,
we find \bea \nonumber
&&\big(\Phi(y_1^+,y_2^+)-\Phi(y_1^+,y_2^-)\big)\star\Delta_{Q'1}=
\Phi\big[y_1^+,x(v+i0)\big]-\Phi\big[y_1^+,x(v-i0)\big] \, . \eea
According to the formula (\ref{basicx}), $x(v)$ has the property
that $|x(v+iy)|<1$ and \mbox{$|x(v-iy)|>1$,} where $v$ is real and
${\rm Im}\, y>0$. Thus, the expression above equals to the jump
discontinuity of the $\Phi$-function through the unit circle. It
is given by
 \bea
\big(\Phi(y_1^+,y_2^+)-\Phi(y_1^+,y_2^-)\big)\star\Delta_{Q'1}=-\Psi(x(v),y_1^+)\,
.\eea Proceeding in the similar manner, we obtain the contribution
of the first line in eq.(\ref{sigtot3}):
 \bea  &&\big(
\Phi(y_1^+,y_2^+)-\Phi(y_1^+,y_2^-)-\Phi(y_1^-,y_2^+)+\Phi(y_1^-,y_2^-)\big)\star\Delta_{Q'1}=\\
&&~~~~~~~~~~~~~~~~~~~~~\qquad\qquad\qquad\qquad\qquad\qquad=
\Psi(x(v),y_1^-)-\Psi(x(v),y_1^+)\, . \nonumber \eea Contribution
of the second line in eq.(\ref{sigtot3}) is computed exactly in
the same fashion as of the first one. One should use this time the
formula of \cite{AFmtba} for the jump discontinuity of the
$\Psi$-function, when its second argument crosses the unit circle.
As a net result, we find \bea  && -{1\ov
2}\left(\Psi(y_1^+,y_2^+)+\Psi(y_1^-,y_2^+)-\Psi(y_1^+,y_2^-)-\Psi(y_1^-,y_2^-)\right)\star\Delta_{Q'1}=
\\ \nonumber
&&~~~~~~\qquad\qquad\qquad\qquad\qquad=-\frac{1}{2i}\log\frac{u-v+\frac{i}{g}Q}{u-v-\frac{i}{g}Q}
-\frac{1}{i}\log i^Q
\frac{\Gamma\big[\frac{Q}{2}-\frac{i}{2}g(u-v)\big]}{\Gamma\big[\frac{Q}{2}+\frac{i}{2}g(u-v)\big]}\,
. \eea

The contribution of the third line in eq.(\ref{sigtot3}) is a bit
more tricky to figure out. The point is that the action of the
second term in the operator $\Delta_{Q'1}$ puts the second
argument of the function $\Psi$ precisely on the cut of the latter
located at $v+\frac{2i}{g}$. Therefore, to proceed, we have to
specify which value of $\Psi$ we utilize -- ether the one on to
the upper edge of the cut $v+\frac{2i}{g}+i0$ or the one on the
lower edge $v+\frac{2i}{g}-i0$. Of course, the prescription
should be fixed in the universal manner for all the
$\Psi$-functions appearing in the third line of
eq.(\ref{sigtot3}).

\smallskip

It turns out, quite remarkably, that the net result does not
depend on which prescription is used. As soon as a choice is made,
the action of $\Delta_{Q'1}$ reduces to evaluation of the jump
discontinuity of $\Psi$ through the cut, which has been already
done in \cite{AFdf}. We spear the details of the computation
presenting only the final result
 \bea && {1\ov
2}\left(\Psi(y_{2}^+,y_1^+)+\Psi(y_{2}^-,y_1^+)-\Psi(y_{2}^+,y_1^-)
-\Psi(y_{2}^-,y_1^-) \right)\star\Delta_{Q'1}=\\ \nonumber &&
~~~~~~~~~~~~~~~~~~~=\Psi(x(v),y_1^+)-\Psi(x(v),y_1^-)+\frac{1}{2i}\log\frac{y_1^+-x(v)}{y_1^+-\frac{1}{x(v)}}
\frac{y_1^--\frac{1}{x(v)}}{y_1^--x(v)}\, . \eea The last term
here can be also represented in the following form
\bea\hspace{-0.5cm}
\frac{1}{2i}\log\frac{y_1^+-x(v)}{y_1^+-\frac{1}{x(v)}}
\frac{y_1^--\frac{1}{x(v)}}{y_1^--x(v)}=
-\frac{1}{i}\log\frac{y_1^+-\frac{1}{x(v)}}{y_1^--\frac{1}{x(v)}}\sqrt{\frac{y_1^-}{y_1^+}}
+\frac{1}{2i}\log\frac{u-v+\frac{i}{g}Q}{u-v-\frac{i}{g}Q}\, .
\eea Finally, contribution of the forth line in eq.(\ref{sigtot3})
is straightforward to find \bea\nonumber && \hspace{-2cm} {1\ov
i}\log\frac{ i^{Q}\,\Gamma\big[Q'-{i\ov
2}g\big(y_1^++\frac{1}{y_1^+}-y_2^+-\frac{1}{y_2^+}\big)\big]} {
i^{Q'}\Gamma\big[Q+{i\ov
2}g\big(y_1^++\frac{1}{y_1^+}-y_2^+-\frac{1}{y_2^+}\big)\big]}{1-
{1\ov y_1^+y_2^-}\ov 1-{1\ov
y_1^-y_2^+}}\sqrt{\frac{y_1^+y_2^-}{y_1^-y_2^+}}\star\Delta_{Q'1}=\\
&&~~~~~~~~~~~~ =\frac{1}{i}\log i^Q
\frac{\Gamma\big[\frac{Q}{2}-\frac{i}{2}g(u-v)\big]}{\Gamma\big[\frac{Q}{2}+\frac{i}{2}g(u-v)\big]}
+\frac{1}{i}\log\frac{y_1^+-\frac{1}{x(v)}}{y_1^--\frac{1}{x(v)}}\sqrt{\frac{y_1^-}{y_1^+}}\,
. \eea Summing up all the contributions, we find zero, {\it i.e.}
$$
\log\check{\Sigma}_Q(u,v)\equiv
\log\Sigma_{QQ'}(y_1,y_2)\star\Delta_{Q'1}=0~~~~{\rm for}~~~v\in
(-2,2)\, .
$$
Now we turn to the second case.

\vskip 0.4cm \noindent \underline{\bf Case II: $|v|>2$.}

\bigskip

In what follows we introduce the concise notation $x\equiv
x(v-i0)$  which represents the (real) value of $x$ on the lower
edge of the cut $]-\infty,-2]\cup[2,\infty[$. The value on the
upper edge is then $x(v+i0)=1/x(v-i0)$. Further, the function $x$
can be conveniently represented as
$$x = {1\ov
2}\big(v+\sqrt{v^2-4}\big)\theta(v-2) + {1\ov
2}\big(v-\sqrt{v^2-4}\big)\theta(-v-2)\, . $$ Evaluation of the
action of $\Delta_{Q'1}$ on $\log\Sigma_{QQ'}(y_1,y_2)$ does not
contain any subtlety and we quote here the corresponding result
\bea \la{Case2}\begin{aligned}
\frac{1}{i}\log\check{\Sigma}_Q(u,v)&= \frac{1}{i}\log\Sigma_{QQ'}(y_1,y_2)\star\Delta_{Q'1}= \\
 &= \Phi(y_1^-,x)-\Phi(y_1^-,{1\ov x})  -
\Phi(y_1^+,x)+\Phi(y_1^+,{1\ov x})\\ &+{1\ov 2}\Big(
\Psi(y_1^-,x)-\Psi(y_1^-,{1\ov x}) +
\Psi(y_1^+,x)-\Psi(y_1^+,{1\ov x})\Big) \\
&+\Psi(x,y_1^+)-\Psi(x,y_1^-)
\,  \\
 &+ \frac{1}{i}\log i^Q
\frac{\Gamma\big[\frac{Q}{2}-\frac{i}{2}g(u-v)\big]}{\Gamma\big[\frac{Q}{2}+\frac{i}{2}g(u-v)\big]}
+\frac{1}{i}\log\frac{y_1^+-{1\ov x}}{y_1^--x}\sqrt{x^2\,
\frac{y_1^-}{y_1^+}}\,
 .
 \end{aligned}
 \eea
Applying the derivative $\frac{1}{2\pi }\frac{d}{du}$ to the last
formula, yields the kernel $\check{K}_{Q}^\Sigma$ for $|v|>2$.

\medskip

It appears that the dressing kernel
$\check{K}_{Q}^\Sigma=\frac{1}{2\pi
i}\frac{d}{du}\log\check{\Sigma}_Q$ has a nice representation in
terms of simpler kernels appearing in the TBA equations. To find
it, we first note that since all the kernels, except for the
dressing one, are defined on the (part of) real $u$-line, it is
natural to transform the integration contours (circles) in the
integrals entering eq.(\ref{Case2}) into the interval $[-2,2]$.
This is easily done by noting that for an arbitrary function
$f(w)$ on a circle, $w=e^{i\theta}$, such that $f(w)=f(1/w)$ one
has \bea &&\int\frac{dw}{2\pi i}\frac{1}{w-x}f(w)=
\int_{-2}^2\frac{dz}{2\pi}\frac{1}{\sqrt{4-z^2}}\frac{2-zx}{x\big[x+\frac{1}{x}-z\big]}f(z)\,
. \eea
 This formula, in conjunction with the identity
$$(1-x(v)^2)/x(v)=-\sqrt{v^2-4}\big[\theta(v-2)-\theta(-v-2)\big]$$
and the following properties of $K_{Qy}$
 \bea K_{Qy}(u,2) = K_{Qy}(u,-2)=0\, ,
\eea allows one to derive the following formula \bea\la{f2}
&&{\pa\ov \pa u}\Big[ \Phi(y_1^-,x)-\Phi(y_1^-,{1\ov x})
-\Phi(y_1^+,x)+\Phi(y_1^+,{1\ov x})\Big]= \\\nonumber
&&~~~~~~~~~~~~=\pm {i\ov \pi}\int_{-2}^2 dt_1 {dt_2\ov
\sqrt{4-t_2^2}}K_{Qy}(u,t_1)\frac{\sqrt{v^2-4}}{t_2-v}
 {d\ov dt_1} \log
\frac{\Gamma\big[1-\frac{i}{2}g(t_1-t_2)\big]}{\Gamma\big[1+\frac{i}{2}g(t_1-t_2)\big]}
\,, \eea where an overall $``+"$ sign is for $v>2$ and $``-"$ for
$v<-2$, respectively.
With the help of the kernel \eqref{KG0}
the formula \eqref{f2} can be written as the double convolution
\bea\la{f3} {1\ov 2\pi}{\pa\ov \pa u}\Big[
\Phi(y_1^-,x)-\Phi(y_1^-,{1\ov x}) -\Phi(y_1^+,x)+\Phi(y_1^-,{1\ov
x})\Big]=-2 K_{Qy}\star K_{\Gamma}^{[2]}\star\check{K}\, ,~~~~ \eea
where both integrations are taken over the interval $[-2,2]$.

\medskip

Now we investigate the second line of eq.(\ref{Case2}). Performing
the same steps as above, we find the following identity
\bea\la{dPsi12vg3} && \Delta\Psi\equiv {1\ov 2}\Big(
\Psi(y_1^-,x)-\Psi(y_1^-,{1\ov x})  +
\Psi(y_1^+,x)-\Psi(y_1^+,{1\ov x})\Big)\\\nonumber && ~~~~~~~~
=\pm{1\ov 2\pi i} \int_{-2}^2  {dt\ov \sqrt{4-t^2}}
\frac{\sqrt{v^2-4}}{t-v}\log \frac{\Gamma\big[1+{Q\ov
2}-\frac{i}{2}g(u-t)\big]}{\Gamma\big[1+{Q\ov
2}+\frac{i}{2}g(u-t)\big]}\frac{\Gamma\big[1-{Q\ov
2}-\frac{i}{2}g(u-t)\big]}{\Gamma\big[1-{Q\ov
2}+\frac{i}{2}g(u-t)\big]}\,, \eea where again an overall $``+"$
sign is for $v>2$ and $``-"$ for $v<-2$, respectively.
One further
finds that
\bea\la{KGQ1}
 {1\ov 2\pi i}{d\ov d u}
  \log\frac{\Gamma\big[1+{Q\ov 2}-\frac{i}{2}g u\big]}{\Gamma\big[1+{Q\ov 2}+\frac{i}{2}gu\big]}
  \frac{\Gamma\big[1-{Q\ov 2}-\frac{i}{2}gu\big]}{\Gamma\big[1-{Q\ov 2}+\frac{i}{2}gu\big]}
   =K_{\Gamma}^{[Q+2]}+K_{\Gamma}^{[Q]}  \,.~~~~ \eea
Furthermore\footnote{Another interesting relation is $K_Q\star
K_{\Gamma}^{[2]}=K_{\Gamma}^{[Q+2]}$, where integration is
performed over the whole real line.}, \bea
K_{\Gamma}^{[Q+2]}=K_{\Gamma}^{[Q]}-K_Q\, ~~~~~\, . \eea
Therefore, the contribution corresponding to the second line takes
the form \bea \frac{1}{2\pi}\frac{\pa}{\pa
u}\Delta\Psi=(2K_{\Gamma}^{[Q]}-K_Q)\star \check{K}\, . \eea

\medskip

For the third line in eq.(\ref{Case2}) we find \bea\la{fp1}
{\pa\ov \pa u}\Big[ \Psi(x,y_1^+) -\Psi(x,y_1^-)\Big]= \mp i\,
\int_{-2}^2 dt \, {\pa\ov \pa t} K_{Qy}(u,t)\log
\frac{\Gamma\big[1+\frac{i}{2}g(v-t)\big]}{\Gamma\big[1-\frac{i}{2}g(v-t)\big]}\,.
\eea where  an overall $``+"$ sign is for $v<-2$ and $``-"$ for
$v>2$, respectively. Integrating by parts, we get \bea\la{fp2}
{1\ov 2\pi}{\pa\ov \pa u}\Big[ \Psi(x,y_1^+) -\Psi(x,y_1^-)\Big]=
- K_{Qy}\star K_{\Gamma}^{[2]}\,,~~~~ \eea

\medskip

Finally, we notice that for $|v|>2$ the following identity is
valid  \bea\la{fp4} \frac{1}{2\pi i}{\pa\ov \pa
u}\log\frac{y_1^+-{1\ov x}}{y_1^--x}\sqrt{x^2\,
\frac{y_1^-}{y_1^+}} =- {1\ov 2}\check{K}_Q(u,v) -  {1\ov
2}K_Q(u-v)\, . \eea  Thus, for the last line in eq.(\ref{Case2})
one gets
\bea\la{fp5} &&{1\ov 2\pi i} {\pa\ov \pa u} \log
\frac{\Gamma\big[{Q\ov 2}-\frac{i}{2}g
(u-v)\big]}{\Gamma\big[{Q\ov 2}+\frac{i}{2}g
(u-v)\big]}\frac{y_1^+-{1\ov x}}{y_1^--x}\sqrt{x^2\,
\frac{y_1^-}{y_1^+}} \nonumber \\
&&\qquad\qquad\qquad\qquad = K_\Gamma^{[Q]}(u-v)-{1\ov
2}\check{K}_Q(u,v)-\frac{1}{2}K_Q(u-v) \,.~~~~~ \eea

\medskip

Combining everything together, we find \bea\la{totsigmacheck}
\check{K}^{\Sigma}_Q=\frac{1}{2\pi i}\frac{\pa}{\pa
u}\check{\Sigma}_Q(u,v)&=&-2K_{Qy}\star
K_{\Gamma}^{[2]}\star\check{K}+(2K_{\Gamma}^{[Q]}-K_Q)\star\check{K}
\\ \nonumber &&~~~~~~~~~~~~~-
K_{Qy}\star
K_{\Gamma}^{[2]}+K_{\Gamma}^{[Q]}-\frac{1}{2}\check{K}_Q-\frac{1}{2}K_{Q}\,
. \eea We stress again that all the convolutions here are taken
from $-2$ to $2$. The kernels involved in the last formula satisfy
a number of magic properties, which lead to a significant
simplification of eq.(\ref{totsigmacheck}). First, one has
\bea\la{sA}
K_Q\star\check{K}=\frac{1}{2}\check{K}_Q-\frac{1}{2}K_Q\, ,
~~~~1\star\check{K}=-\frac{1}{2}\, . \eea These relations allow one
to find \bea\la{sB}
K_{\Gamma}^{[Q]}\star\check{K}=\frac{1}{2}\check{K}_Q-\frac{1}{2}K_{\Gamma}^{[Q]}+\frac{1}{2}\sum_{n=1}^{\infty}\check{K}_{2n+Q}\,
. \eea Specifying the last expression for $Q=2$,
 one gets
\bea\la{sC}
K_{\Gamma}^{[2]}\star\check{K}=\frac{1}{2}\check{K}_2-\frac{1}{2}K_{\Gamma}^{[2]}+\frac{1}{2}\sum_{n=1}^{\infty}\check{K}_{2n+2}=
-\frac{1}{2}K_{\Gamma}^{[2]}+\frac{1}{2}\sum_{n=1}^{\infty}\check{K}_{2n}\,
. \eea
Applying the identities (\ref{sA})-(\ref{sC}) in the first
line of eq.(\ref{totsigmacheck}), we find the following simple
result
\bea \check{K}_Q^\Sigma(u,v) =- K_{Qy}\star
\sum_{n=1}^\infty \check{K}_{2n} + \sum_{n=1}^\infty
\check{K}_{2n+Q}\,.~~~~~~~ \eea We note that for numerical
computations, the fastest algorithm consists in replacing the
infinite sums in the last formula by their integral
representation
 \bea
 \la{IQ}
 \check{I}_Q&=&\sum_{n=1}^\infty
\check{K}_{2n+Q}(u,v)=K_\Gamma^{[Q+2]}(u-v)+2\int_{-2}^2 {\rm d}t
\, K_\Gamma^{[Q+2]}(u-t)\check{K}(t,v) \,.~~~~~ \eea


\end{document}